\documentclass[fleqn,usenatbib]{mnras}

\usepackage{newtxtext,newtxmath}

\usepackage[T1]{fontenc}
\usepackage{ae,aecompl}

\usepackage{graphicx}	
\usepackage{amsmath}	
\usepackage{amssymb}	
\usepackage{subfig} 	
\usepackage{xcolor}		

\newcommand{\e}[1]{\times 10^{#1}}

\title[Jovian Trojan Stability]{Stability of Jovian Trojans and their collisional families}

\author[T. R. Holt et al.]{
Timothy R. Holt,$^{1,2}$\thanks{E-mail: timothy.holt@usq.edu.au (TRH)}
David Nesvorn{\'y},$^{2}$
Jonathan Horner,$^{1}$
Rachel King,$^{1}$
\newauthor
Raphael Marschall,$^{2}$
Melissa Kamrowski, $^{3}$
Brad Carter,$^{1}$
\newauthor
Leigh Brookshaw,$^{1}$
and Christopher Tylor,$^{1}$
\\
$^{1}$University of Southern Queensland, Centre for Astrophysics, Toowoomba, Queensland 4350, Australia\\
$^{2}$Southwest Research Institute, Department of Space Studies, Boulder, CO. USA.\\
$^{3}$University of Minnesota, Morris, MN. USA.
}

\date{Accepted XXX. Received YYY; in original form ZZZ}

\pubyear{2019}
\begin{document}
\label{firstpage}
\pagerange{\pageref{firstpage}--\pageref{lastpage}}
\maketitle

\begin{abstract}
The Jovian Trojans are two swarms of objects located around the L$_4$ and L$_5$ Lagrange points. The population is thought to have been captured by Jupiter during the Solar system's youth. Within the swarms, six collisional families have been identified in previous work, with four in the L$_4$ swarm, and two in the L$_5$.
Our aim is to investigate the stability of the two Trojan swarms, with a particular focus on these collisional families. 
We find that the members of Trojan swarms escape the population at a linear rate, with the primordial L$_4$ (23.35\% escape) and L$_5$ (24.89\% escape) population sizes likely 1.31 and 1.35 times larger than today.
Given that the escape rates were approximately equal between the two Trojan swarms, our results do not explain the observed asymmetry between the two groups, suggesting that the numerical differences are primordial in nature, supporting previous studies.
Upon leaving the Trojan population, the escaped objects move onto orbits that resemble those of the Centaur and short-period comet populations.
Within the Trojan collisional families, the 1996 RJ and 2001 UV$_{209}$ families are found to be dynamically stable over the lifetime of the Solar system, whilst the Hektor, Arkesilos and Ennomos families exhibit various degrees of instability.
The larger Eurybates family shows 18.81\% of simulated members escaping the Trojan population. Unlike the L4 swarm, the escape rate from the Eurybates family is found to increase as a function of time, allowing an age estimation of approximately $1.045\pm 0.364\e9$ years.

\end{abstract}




\section{Introduction}
\label{Sec:Intro}


The Jovian Trojans are a population of small Solar system bodies comprising two swarms located around the leading (L$_4$) and trailing (L$_5$) Lagrange points of Jupiter. The larger and better known members of the Trojan swarms are named after the characters of the epic Greek poems that detail the Trojan war, The Iliad and The Odyssey \citep{HomerIliad}. 

The Jovian Trojans were discovered in the early 20th Century, with the first, \citep[588 Achilles,][]{Wolf1907588Achillies} being quickly followed by 617 Patroclus, 624 Hektor and 659 Nestor \citep{Heinrich1907617Patroclus,Stromgren1908624Hektor,Ebell1909659Nestor,Kopff1909HekorNestor}. These objects were the first confirmation of a stable solution to the restricted three-body problem that had been proposed over a century earlier by \citet{Lagrange1772essai}. 

At the time of writing, approximately 7200 objects have been discovered around the Lagrange points of Jupiter\footnote{Taken from the JPL HORIZONS Solar System Dynamics Database \url{https://ssd.jpl.nasa.gov/} \citep{Giorgini1996JPLSSdatabase}, on 13th November, 2019.}, a number that is destined to rise still further in the coming years, as a result of the Rubin Observatory Legacy Survey of Space and Time (LSST), scheduled for first light in 2021 \citep{Schwamb2018LSSTTrojans}. Interestingly, the known Trojans are not evenly distributed between the two Trojan swarms. Instead, there is a marked asymmetry, with the leading L$_4$ swarm containing approximately 1.89 times the number of objects than the L$_5$ swarm. A number of studies have considered this asymmetry, and have found it to be robust, a real feature of the population, rather than being the result of observational biases \citep{Jewitt2000JovTrojan, Nakamura2008TrojanSurfaceDenSizeEst,Yoshida2008JovTrojanSize, Vinogradova2015JupTrojanMass}.

\begin{figure}
	\includegraphics[width=\columnwidth]{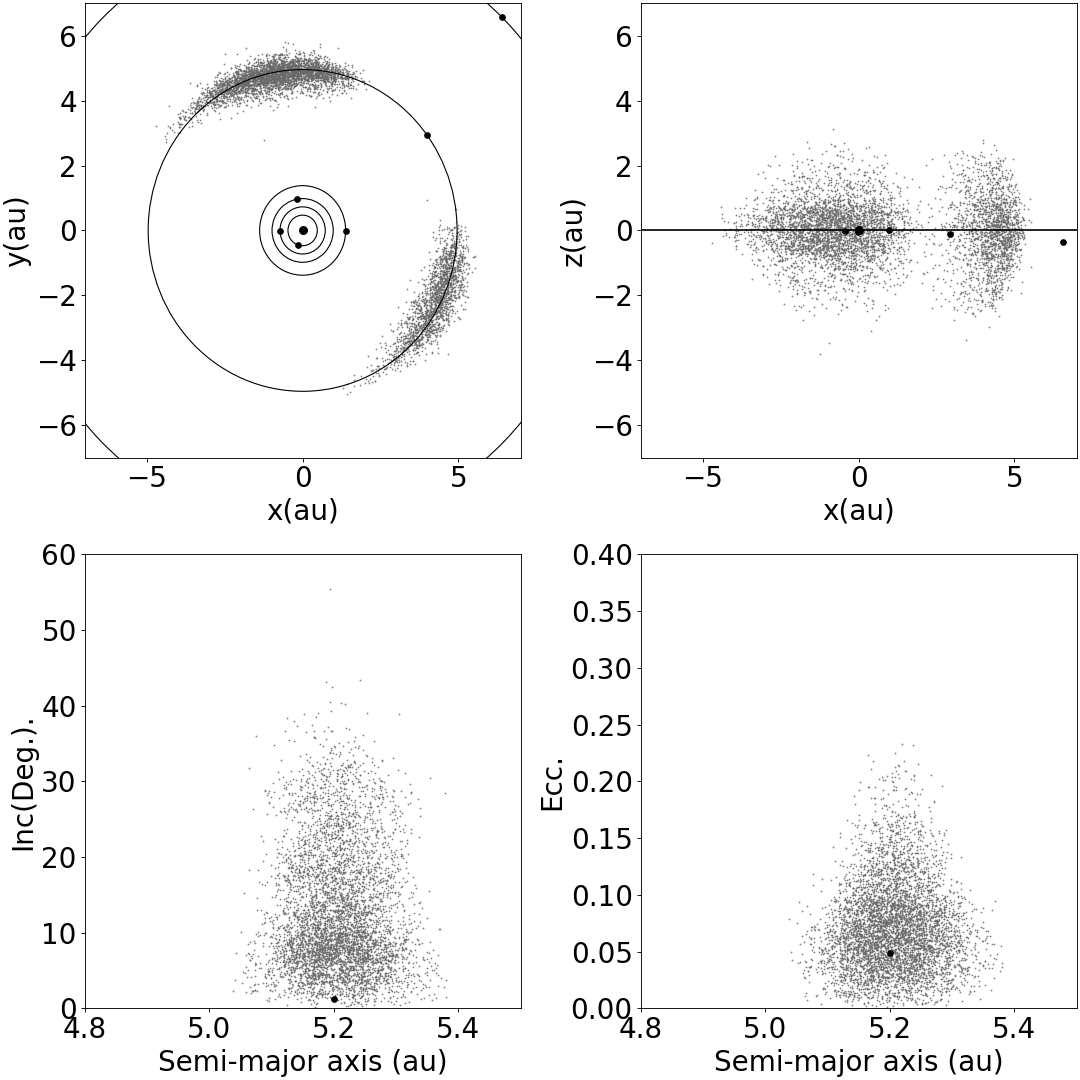}
    \caption{Distribution of 5553 Jovian Trojans for which proper elements have been generated \citep{Knezevic2017AstDysTrojans}. The top figures indicate the positions of the Trojans relative to the planets on 01-01-2000 00:00 in a face-on (xy; left) and edge-on (xz; right) orientation, in the ecliptic reference system. Bottom figures show the Trojans in osculating inclination (Inc), eccentricity (Ecc) and semi-major axis space. Larger black dots indicate planets, with Jupiter being shown in the bottom diagrams. Data from NASA HORIZONS, as of 19th Aug. 2019.}
    \label{Fig:TrojansFig}
\end{figure}

Although more than 7200 objects have been found in the region surrounding the Jovian Lagrange points, many of those objects may be temporarily captured objects, rather than permanent members of the Trojan population. Whilst the `true' Trojans move on stable orbits that keep them librating around the L$_4$ and L$_5$ Lagrange points on billion year timescales \citep[e.g.][]{EmeryAsteroidsIVJupTrojan}, temporarily captured objects would be expected to escape from the Trojan swarms on timescales of thousands or tens of thousands of years. To confirm that a given object is truly a member of the Trojan population requires confirmation that the object's proper orbital elements \citep{Milani1992ProperElements} are stable, and that the object is truly trapped in 1:1 resonance with Jupiter. Simulations spanning more than $1\e6$ years and transformation using Fourier transform analysis \citep{Sidlichovsky1996FMFT, Beauge2001TrojanFamilies, Broz2011EurybatesFamily} are used to devolve the osculations of potential Trojans, to determine whether or not their orbits are truly resonant. The database of those objects for which such analysis has been carried out can therefore be considered a set of contemporary stable Jovian Trojans, and includes 5553 numbered and multi-oppositional objects \citep{Knezevic2017AstDysTrojans}. Fig. \ref{Fig:TrojansFig} shows the current known configuration of the Jovian Trojan population.

\begin{figure}
	\includegraphics[width=\columnwidth]{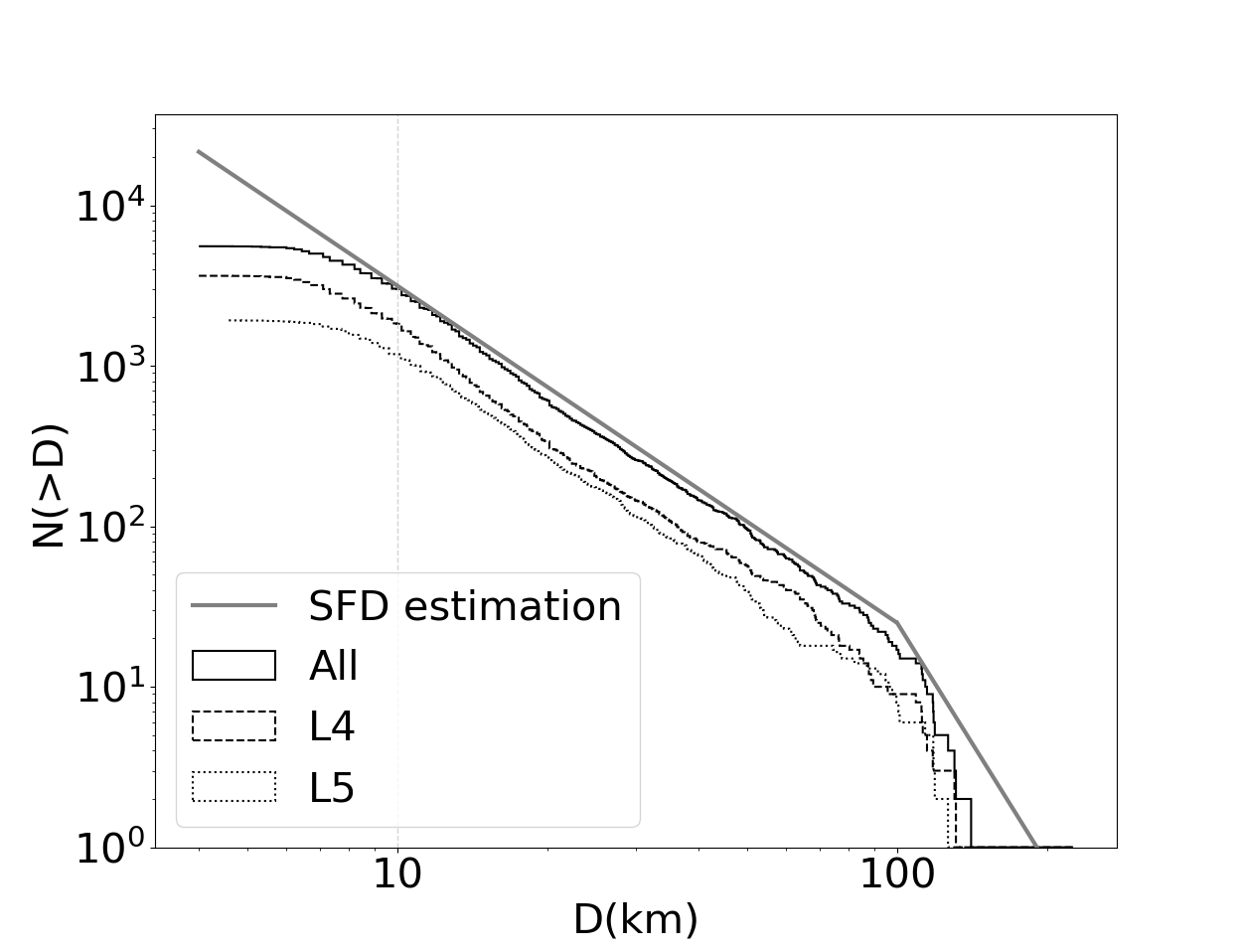}
    \caption{Cumulative size-frequency distribution of the Jovian Trojans. The solid line shows the distribution for the population as a whole, whilst the long-dash line shows the distribution among members of the leading L$_4$ swarm, and the dotted line shows the distribution for the trailing L$_5$ swarm. Data from NASA HORIZONS, as of 19th Aug. 2019. Vertical grey, dashed line indicates observational completeness \citep{EmeryAsteroidsIVJupTrojan}. The grey line shows the estimated complete size distribution \citep{Nesvorny2018SSDynam}.}
    \label{Fig:TrojansSFD}
\end{figure}

In order to asses the observational completeness of the Trojan population, an examination of their size distribution is needed. The observed population of Jovian Trojans ranges in diameter from the largest, 624 Hektor, at $\sim$ 250 km \citep{Marchis2014Hektor}, down to objects several kilometres across \citep{EmeryAsteroidsIVJupTrojan}. The size-frequency distribution for these objects is generally considered to be observationally complete to approximately 10km in size \citep{EmeryAsteroidsIVJupTrojan, Grav2011JupTrojanWISEPrelim}, as shown in Fig. \ref{Fig:TrojansSFD}. The power law that best describes this size distribution is similar to that of the collisionally evolved Asteroid belt \citep{Bottke2005SFDCollisions}. From this it has been inferred that the Jovian Trojan population could contain as many as a million objects greater than 1km in diameter \citep{Jewitt2000JovTrojan, Yoshida2008JovTrojanSize, Yoshida2017JupTrojanSize}, though there are also indications that these may be  optimistic estimates that grossly overestimate the true situation \citep[e.g.][]{Nakamura2008TrojanSurfaceDenSizeEst}. 

\subsection{The dynamics and origins of the Jovian Trojans}
Due to their stability, it is thought that the Jovian Trojans date back to the early Solar system \citep[e.g.][]{EmeryAsteroidsIVJupTrojan,Nesvorny2018SSDynam}. Attempts to ascertain the origins of the Jovian Trojans need to explain their unique dynamical situation. As can be seen in Fig.  \ref{Fig:TrojansFig}, the population is dynamically `warm', occupying two broad tori around the Lagrange points, with high orbital inclinations and eccentricities. An in-situ formation would be expected to produce a `cold' disk, with low orbital eccentricities and inclinations, reflective of the primordial protoplanetary disk. The mismatch between the observed population and the distribution that would be expected from in-situ formation has led to the conclusion that the Jovian Trojans most likely did not form in their current orbits, but were in fact captured early in the Solar system's history \citep[e.g.][]{Morbidelli2005TrojanCapture,Lykawka2010TrojanCapMigration,Nesvorny2013TrojanCaptureJJ,Pirani2019PlanetMigrationSSB}.

One explanation for the observed orbital distribution of the Jovian Trojans comes in the form of the `Nice' Model. This model invokes a period of chaotic disruption in the outer Solar system to explain the origin of the Late Heavy Bombardment \citep{Tsiganis2005NICEplanets,Morbidelli2010NiceReview, Levison2011Nice,Nesvorny2012JumpingJupiter,Deienno2017PlanetInstability,Nesvorny2018SSDynam}, during which the Trojans were trapped in their current orbits from a population of dynamically unstable objects that were being scattered through the outer Solar system \citep{Morbidelli2005TrojanCapture,Lykawka2010TrojanCapMigration,Nesvorny2013TrojanCaptureJJ}. A recent attempt to explain the observed asymmetry, which is not explained by the `Nice' model,  proposes an alternative, that the Trojans were captured from the same region of the disc as Jupiter, and were transported during the planet's proposed inward migration \citep{Pirani2019PlanetMigrationSSB}. In an update to this in-situ transport model, \citet{Pirani2019TrojanInc} explains the inclinations by invoking mixing in the Jovian feeding region. These two competing theories for the origins of the Trojans highlight the importance of the population in our understanding of the early Solar system.

Previous long term simulations of the Jovian Trojans \citep{Levison1997JupTrojanEvol,Tsiganis2005ChaosJupTrojans, DiSisto2014JupTrojanModels,DiSisto2019TrojanEscapes} have indicated that at least some of the members of both the L$_4$ and L$_5$ swarms are actually temporary captures, and will escape from the Trojan swarms on timescales of $\sim 1\e{6}$ years. The estimated fraction of Trojans that will escape the population on these timescales varies somewhat between these studies, with \citet{Levison1997JupTrojanEvol} proposing an escape rate of $\sim$ 12\%, and \citet{Tsiganis2005ChaosJupTrojans} estimating 17\%. More recent works, by \citet{DiSisto2014JupTrojanModels,DiSisto2019TrojanEscapes}, suggest a still higher escape rate, at 23\% for the L$_4$ and 28\% for the L$_5$ swarm. To some extent, the disparity among these results can be explained by the growth in the known Trojan population that occurred between one study and the next. \citet{Levison1997JupTrojanEvol} considered a sample of only 178 numbered objects. In contrast, \citet{Tsiganis2005ChaosJupTrojans} studied 246 numbered objects. The 2972 numbered Trojans that were simulated by \citet{DiSisto2014JupTrojanModels,DiSisto2019TrojanEscapes} make it the largest previous study. 

To further complicate the picture, detailed modeling of (1173) Anchises \citep{Horner2012AnchisesThermDynam} has shown that at least some of the unstable Jovian Trojans could still be primordial in nature. Indeed, that work, along with other studies in stability \citep{Levison1997JupTrojanEvol, Nesvorny2002Trojans, Tsiganis2005ChaosJupTrojans, DiSisto2014JupTrojanModels, DiSisto2019TrojanEscapes} suggests that the original population of Jovian Trojans was larger than that observed today, and that it likely included objects with a range of stabilites. (1173) Anchises is stable on timescales of hundreds of millions of years, and so might well be a representative of a once larger population of such objects, which have slowly escaped from the Trojan population since their formation. Following a similar argument, \citet{Lykawka2010TrojanCapMigration} propose a link between the Centaur population and the Jovian Trojans that escape, though this is disputed by \citet{Jewitt2018TrojanColor} due to differences in the colour distributions of the two populations. \citet{Wong2016ColorJupTrojan} also use the observed colours of members of the Jovian Trojan population to propose a hypothesis for a common origin between the Trojans and the Edgeworth-Kuiper Belt objects. Such an origin is a good fit with the results of dynamical models that invoke an instability in the outer Solar system as the origin of the Jovian Trojans, in which the Jovian Trojans are captured from a similar source region to the Edgeworth-Kuiper Belt objects \citep{Morbidelli2005TrojanCapture,Nesvorny2013TrojanCaptureJJ}.

\subsection{Collisional Families amongst the Jovian Trojans}
Elsewhere in the Solar system, other evolved populations contain dynamical families, the results of the collisional disruption of large parent bodies. Such collisional families have been identified in the asteroid main belt \citep[see][]{Hirayama1918AsteroidFam, Gradie1979AstFamReview, Zappala1984AsteroidCollision, Knezevic2003AstDys, Carruba2013AsteroidFamilies, Milani2014AsteroidFamilies,  Nesvorny2015AsteroidFamsAIV, Milani2017AsteroidFamAges},the Hilda \citep{Broz20081stOrderResAst} and Hungaria \citep{Warner2009HungariaFam,Milani2010HungariaDynam} populations, the irregular satellites of the giant planets \citep{Nesvorny2003IrrSatEvol, Sheppard2003IrrSatNature, Grav2003IrregSatPhoto, Nesvorny2004IrrSatFamilyOrigin, Grav2007IrregSatCol, Jewitt2007IrregularSats, Turrini2008IrregularSatsSaturn, Turrini2009IrregularSatsSaturn, Bottke2010IrrSatSFD, Holt2018JovSatSatsClad} and the Haumea family in the Edgeworth-Kuiper belt \citep{Brown2007HaumeaFamily, Levison2008HaumeaFamOrigin, delaFuente2018OSSFams}. The traditional methodology for identifying these families in small body populations was developed by \citet{Zappala1990HierarchicalClustering1, Zappala1994HierarchicalClustering2} and is known as the Hierarchical Clustering Method (HCM), and utilises distances in semi-major axis, eccentricity and inclination parameter space to identify family members. 

Historically, studies that attempted to identify such collisional families amongst the Jovian Trojans were limited by the number of objects that had been discovered at that time \citep{Milani1993JovianTrojFamilies}. Additionally, as the Jovian Trojans librate around the Lagrange points,the calculation of proper elements used in family identification is problematic \citep{EmeryAsteroidsIVJupTrojan}. For that reason, \citet{Beauge2001TrojanFamilies} used transformed proper elements to account for the librations present in the Jovian Trojan dynamics. As the number of known Jovian Trojans increased, additional dynamical clusters have been identified \citep[e.g.][]{Roig2008JovTrojanTaxon, DeLuise2010Eurybates, Broz2011EurybatesFamily, Vinogradova2015TrjoanFamilies, Nesvorny2015AsteroidFamsAIV, Rozehnal2016HektorTaxon}. \citet{Rozehnal2016HektorTaxon} offer an expansion to the HCM developed by \citet{Zappala1990HierarchicalClustering1}. This new `randombox' method uses Monte-Carlo simulations to determine the probability that the identified clusters are random in parameter space. Canonically, six collisional families, four in the L$_4$ swarm and two in the L$_5$, are now considered valid in the Jovian Trojan population \citep{Nesvorny2015AsteroidFamsAIV}. Independent HCM analysis undertaken by \citet{Vinogradova2015TrjoanFamilies} has confirmed the four L$_4$ families, though they dispute the validity of the L$_5$ families. See Table \ref{tab:Families} for details on the families we consider in this work. 

\begin{table}
\centering
\caption{Identified collisional families in the Jovian Trojan swarms, after \citep{Nesvorny2015AsteroidFamsAIV}. FIN: Family identification number, used throughout this manuscript; $n$: Number of family members; $D_{LM}$: Diameter of the largest member; Tax.: Identified taxonomic type \citep{Bus2002AsteroidTax, Grav2012JupTrojanWISE}.}
\label{tab:Families}
\resizebox{\columnwidth}{!}{
\begin{tabular}{lcccc}
\hline
Family         & FIN & $n$ & $D_{LM}$ (km) & Tax. \\
\hline
L$_4$        &     &     &     &      \\
Hektor     & 1 & 12  & 225      & D    \\
Eurybates &2  & 218 & 63.88    & C/P   \\
1996 RJ &3    & 7   & 68.03    & -    \\
Arkesilaos &4 & 37  & 20.37    & -    \\
\hline
L$_5$         &    &     &          &      \\
Ennomos &5    & 30  & 91.43    & -    \\
2001 UV$_{209}$ &6 & 13  & 16.25    & -   
\end{tabular}%
}
\end{table}

Early imaging surveys suggest that there is a spectral commonality within the dynamical families \citep{Fornasier2007VisSpecTrojans} in the Jovian Trojans. More recent observational data has brought this into question \citep{Roig2008JovTrojanTaxon}, with a heterogeneity being seen in some unconfirmed families from the Sloan Digital Sky Survey (SDSS) colors. The confirmed Eurybytes and Hektor families, however, show a distinctive colour separation from the rest of the population \citep{Roig2008JovTrojanTaxon, Broz2011EurybatesFamily, Rozehnal2016HektorTaxon}. \citet{Vinogradova2015TrjoanFamilies} also make comments on the taxonomy of the L4 families, based on SDSS taxonomy \citep{Carvano2010SDSSAstTax}. In these studies, the Eurybates family is found to consist mainly of C-types, and the Hektor family mostly D-types, under the Bus-Demeo taxonomy \citep{Bus2002AsteroidTax, DeMeo2009AsteroidTax}.

Unlike collisional families in the asteroid belt, the determination of ages for the Trojan families remains elusive. Currently there are two general methods used to determine family ages \citep{Nesvorny2015AsteroidFamsAIV}. The first involves reverse integration $n$-body simulations of the identified family. A relatively young family, such as the Karin family \citep{Nesvorny2002AsteroidBreakup}, would show convergence in both longitude of ascending node and argument of pericentre as those simulations approach the time of the family's birth. However, such simulations are not able to provide firm constraints on the ages of older families, as a result of the chaotic diffusion experienced by the members of those families over time. Once such diffusion has had sufficient time to act, reverse integration of family members will fail to show such convergence. A variation on this uses synthetic families to estimate the collisional family age \citep{Milani1994VeritasAge, Nesvorny2002FloraFamily}. Some synthetic simulations by \citet{Broz2011EurybatesFamily} and \citet{Rozehnal2016HektorTaxon} have calculated the age of the Hektor, Eurybates and Ennomos families in the Trojan population, though these have relatively large, Gigayear ranges. In order to circumvent some of these issues, a second method of family age estimation was developed. This method relies on the modelling of asteroidal Yarkovsky drift \citep{Vokroughlicky2006YarkovskyAsteroidFamilies, Spoto2015AsteroidFamAges, Bolin2017YarkovskiAsteroidFam}. The technique takes advantage of the fact that any collisional family will contain a large number of different sized objects, which would be expected to experience Yarkovsky drift \citep{Bottke2006YarkovskyAsteroids} at different rates. As a result, when the members of a collisional family are plotted in size, or its proxy absolute magnitude, vs orbital semi-major axis, they will form a characteristic 'V shape' \citep{Vokroughlicky2006YarkovskyAsteroidFamilies,Spoto2015AsteroidFamAges,Paolicchi2019YORPeye}. The slope of the 'V' can then be used to estimate the age of the family. Using this method, a $4\e{9}$ year old meta-family has been identified in the asteroid belt \citep{Delbo2017PrimodialAstFam}. This method has been attempted with the Eurybates family \citep{Milani2017AsteroidFamAges}, though due to the negligible Yarkovsky effect experienced by the Jovian Trojans, the age is unreasonably estimated at $1.4\e{10}$ years. This indicates that the method is inappropriate for age estimation of collisional families in the Jovian Trojan swarms. 

\subsection{This work}
In this work, we utilise $n$-body simulations of the known Jovian Trojan population to consider the stability of previously identified collisional families \citep{Nesvorny2015AsteroidFamsAIV}. This work considers 5553 numbered and multi-oppositional objects, a sample nearly double that of the previous largest study, \citet{DiSisto2014JupTrojanModels,DiSisto2019TrojanEscapes}, who considered 2972 numbered objects. By simulating the whole known population, we can include all identified collisional family members in the study. We divide this work into the following sections. Section \ref{Sec:Methods} describes the methodology of the $n-$body simulations used as the basis for this work. We discuss the L$_4$ and L$_5$ swarms in section \ref{Sec:L4L5Swarms}. In section \ref{SubSec:EscapeAnalysis} we use our simulations to study the rate at which objects escape from the Trojans, and discuss the implications of our results for the original size of the population, including the L$_4$/L$_5$ asymmetry and formation scenarios. We consider the stability of the collisional families in section \ref{Sec:CollisionalFamilies}, with a particular focus on the large Eurybates family in \ref{SubSec:Eurybates}. Concluding remarks are presented in section \ref{Sec:Conclusions}.

\section{Methods}
\label{Sec:Methods}
We selected the Jovian Trojan population for our simulations based on several criteria. An initial dataset was obtained from the JPL Small-Body Database \citep{Giorgini1996JPLSSdatabase} by searching for and selecting all objects with orbital semi-major axes between 4.6 au and 5.5 au and an orbital eccentricity less than 0.3. This process yielded an initial selection of 7202 objects, obtained on 17th April, 2018. The ephemeris were retrieved from the NASA HORIZONS database \citep{Giorgini1996JPLSSdatabase} for all objects using an initial time point of A.D. 2000-Jan-01 00:00:00.0000. We then filtered our sample to discard temporarily captured objects by limited selection to those objects present in the AstDys proper element database \citep{Knezevic2017AstDysTrojans}. Since objects in this list require the completion of simulations spanning $1\e6$ years to generate the proper elements of their orbits \citep{Knezevic2017AstDysTrojans}, this set can be considered initially stable objects. Once our sample was filtered in this way, we were left with a total of 5553 nominally `stable' Trojans for this study, including 4780 numbered and 773 multioppositional objects.

In order to investigate the long-term dynamical evolution of the Jovian Trojan population, we carried out a suite of $n$-body integrations using the WFAST symplectic integrator within the {\textit{REBOUND}} $n$-body dynamics package \citep{Rein2012REBOUND,Rein2015WHFAST}. Eight clones of each reference Trojan were created, distributed across the $\pm 1\sigma$ positional uncertainties from the HORIZONS database \citep{Giorgini1996JPLSSdatabase}. These eight $1\sigma$ clones were generated at the vertices of a cuboid in x-y-z space, with the reference particle in the center. Therefore, in this work we followed the evolution of a total of 49,977 collisionless, massless test particles in our simulations, nine particles for each of the 5553 Trojans. Our integrations modelled the evolution of our test particle swarms under the gravitational influence of the Sun and the four giant planets (Jupiter, Saturn, Uranus and Neptune). Each individual simulation thus consisted of the Sun, four giant planets, the initial HORIZONS reference particle and the eight $1\sigma$ clones, with ephemeris in Solar system barycentric coordinates. All simulations were conducted on the University of Southern Queensland's High Performance Computing Cluster, Fawkes. We ran each simulation forward for $4.5\e{9}$ years, with an integration timestep of 0.3954 years, 1/30th of the orbital period of Jupiter \citep{Barnes2004StabilityPlanets}. The orbital elements of every test particle were recorded every $1\e{5}$ years.

The Yarkovsky effect is a non-gravitational force that can act on small bodies \citep{Bottke2006YarkovskyAsteroids}. The effect involves the asymmetric thermal radiation of photons from an object, which imparts a thrust on the object in question. This thrust will gradually change the semi-major axis of a body, with the scale and direction of the induced drift dependent on the thermal properties, axis of rotation and size of the object \citep{Broz2005YarkovUnstabAst, Bottke2006YarkovskyAsteroids}. In the case of the Jovian Trojans, simulations of hypothetical objects have indicated that at small sizes (<1 km), the Yarkovsky effect could impact the stability of the objects \citep{Wang2017YarkovskyTrojanModel, Hellmich2019TrojanYarkovski}. As we are simulating known Jovian Trojans, the majority of the objects are greater than several kilometres in size \citep{EmeryAsteroidsIVJupTrojan}, and have unknown or highly uncertain thermal properties \citep{Slyusarev2014JupiterTrojens, Sharkey2019TrojanIR}. For these reasons, we have not included the Yarkovsky effect in our simulations.

\section{Escapes from the L4 and L5 swarms}
\label{Sec:L4L5Swarms}
In each of our simulations, we track the position of a particle, and record the time it escapes the Jovian Trojan population. A database of the escape times of each particles is presented in the online supplementary material. We define these escapes as occurring once the test particle obtains an osculating semi-major axis of less than 4.6 au or greater than 5.5 au. In Table \ref{Tab:Escapes}, we present the results of our simulations, showing the fraction of the total population that escaped from the Trojan population during our simulations. As part of our calculations, we include the volume of the object, as a proxy for mass. The density is only known for a single C-type Trojan, (617) Patroclus \citep{Marchis2006DensityPatroclus}. With the diversity of taxonomic types seen in even a small number of classified Trojans \citep{Carvano2010SDSSAstTax, Grav2012JupTrojanWISE, DeMeo2013SDSSTaxonomy}, using mass instead of volume could further propagate errors. The volumes were calculated from diameters in the HORIZONS database to a assumed sphere. Where diameters were unavailable, due to no recorded albedo, we made an estimate based on the $H$ magnitude and mean geometric albedo (from NASA HORIZONS) of each Jovian Trojan swarm, following the methodology of \citet{Harris1997AstAlbDiameter}. We use separate geometric albedos for the L4 (0.076) and L5 (0.071) swarms, as they are significantly different \citep{Romanishin2018AlbedoCenJTH}, though close to the mean geonemtric albedo (0.07) identified by \citet{Grav2011JupTrojanWISEPrelim, Grav2012JupTrojanWISE}. There may be a size dependency on the albedos in the Trojan population \citep{Grav2011JupTrojanWISEPrelim,Grav2012JupTrojanWISE,Fernandez2009AlbedosmallTrojan}, though only a relatively small number of objects have been studied in this way. In choosing to use consistent albedos, there may be some discrepancies between this work and future studies, as more robust albedos, diameters and shape models are presented. We note that the observed L$_4$/L$_5$ asymmetry is lower when volume is considered (L$_4$ 1.56 larger), than simply considering the number of known objects (L$_4$ 1.89 larger).

\begin{table*}
	\caption{Escape percentages of Jovian Trojan swarm members. Column headings: $n$: Number of real Trojan members considered in the simulations; $n_{test}$: number of test particles simulated (eight clones, plus initial reference particle); $f_{EscR}$: numerical percentage of reference particles that escape; $f_{VEscR}$: volumetric percentage of reference particles that escape; $f_{EscP}$: numerical percentage Trojan particle pool, Reference and eight $1\sigma$ clones, that escape; $f_{VEscP}$: volumetric percentage Trojan particle pool, Reference and eight $1\sigma$ clones, that escape; $f_{Esc9C}$: numerical percentage Trojans where all nine particles escape; $f_{VEsc9C}$: volumetric percentage of Trojans where all nine particles escape; $>10kmf_{EscP}$: numerical percentage of Trojan particle pool greater than 10km that escape; $>10kmf_{VEscP}$: volumetric percentage of Trojan particle pool greater than 10km that escape}
	\label{Tab:Escapes}

\begin{tabular}{lcccccccccc}
\hline
      & $n$  & $n_{test}$ & $f_{EscR}$ & $f_{VEscR}$ & $f_{EscP}$ & $f_{VEscP}$ & $f_{Esc9C}$ & $f_{VEsc9C}$ & $>10kmf_{EscP}$ & $>10kmf_{VEscP}$\\ \hline
L$_4$    & 3634 & 32706 & 22.23\%    & 22.97\%     & 23.19\%     & 23.35\%      & 5.01\%      & 7.36\%     & 23.28\%     & 23.37\%       \\
L$_5$    & 1919 & 17271 & 24.80\%    & 32.22\%     & 24.89\%     & 24.89\%      & 5.04\%      & 6.07\%     & 24.27\%     & 24.88\%       \\      
\hline
Total & 5553 & 49977 & 23.12\%    & 26.58\%     & 23.77\%     & 23.95\%      & 5.02\%      & 6.56\%     & 23.67\%     & 23.96\%                    
\end{tabular}%
\end{table*}

The escape percentages of our reference particles are larger than the 12\% seen by \citet{Levison1997JupTrojanEvol}. In order to investigate this discrepancy, we consider the instability of the subset of the 178 Jovian Trojans known at the time of \citet{Levison1997JupTrojanEvol}. Using our simulations, we find an reference particle escape rate of 15\%, consistent with \citet{Levison1997JupTrojanEvol} and similar to the 17\% found by \citet{Tsiganis2005ChaosJupTrojans}. \citet{DiSisto2014JupTrojanModels, DiSisto2019TrojanEscapes} considered the 2972 numbered Trojans known at that time, and found escape rates of 23\% and 28.3\% for the L$_4$ and L$_5$ swarms respectively. The \citet{DiSisto2014JupTrojanModels, DiSisto2019TrojanEscapes} results are closer to our escape rates for the reference particles, and the L$_4$ particle pool escapees. The escape percentages in the L$_5$ clone pool are lower in our simulations, closer to that of the L$_4$ swarm and the population as a whole. 

The similar ratios in escape percentages between the two swarms confirm the findings of others \citep{Nesvorny2002SaturnTrojanHypothetical, Tsiganis2005ChaosJupTrojans, Nesvorny2013TrojanCaptureJJ, DiSisto2014JupTrojanModels, DiSisto2019TrojanEscapes}, who argued that the observed Jovian Trojan swarm asymmetry can not be the result of differences in the escape rate between the two Trojan swarms. The difference is therefore more likely due to differences in the number of objects that were initially captured to the swarms.

At first glance, the escape volume differences between the two swarms, shown in Table \ref{Tab:Escapes}, could account for the asymmetry, particularly in terms of the reference particles ($f_{VEscR}$ in Table \ref{Tab:Escapes}). This can be explained by the escape of several large (<100km diameter) reference objects. In the L$_4$ swarm, the reference particles of (1437) Diomedes and (659) Nestor escape the Trojan population. The reference particles of (3451) Mentor, (1867) Deiphobus, and (884) Priamus, in the L$_5$ swarm also escape. (3451) Mentor and (659) Nestor are classified as X-Type \citep{Tholen1984Taxonomy, Bus2002AsteroidTax}. Once the $1\sigma$ clones are taken into account, $f_{VEscP}$ in Table \ref{Tab:Escapes}, this escape asymmetry in the volume is negated, resulting in near identical escape rates for the L$_4$ and L$_5$ swarms. This volumetric escape fraction ($f_{VEscP}$ in Table \ref{Tab:Escapes}) is very similar to the numerical escape fraction ($f_{EscP}$ in Table \ref{Tab:Escapes}) for the population and in each of the swarms. In order to further investigate the volumetric escapes, we can limit our selection to just objects for which the population can be considered to be observationally complete, those larger than 10km \citep{EmeryAsteroidsIVJupTrojan}. This reduces the numerical size of the population to 3003. When we repeat the analysis, the percentage of particles that escape only changes by fractions of a percent in the population, as well as each swarm, see $>10kmf_{EscP}$ and  $>10kmf_{VEscP}$ in Table \ref{Tab:Escapes}. This additional analysis supports the hypothesis that the observed asymmetry between the swarms is due to implantation, rather than any volumetric differences.

We generate a conservative subset of the escape population, one where all nine particles of a given object escape. In this subset, $f_{Esc9C}$ and $f_{VEsc9C}$ in Table \ref{Tab:Escapes}, escape percentages are much lower. These escapes represent the minimal set of escapes and show that the majority of the escaping population are statistically boarder-line. Those objects where all nine particles escape are deep into the parameter space identified as unstable by \citet{Levison1997JupTrojanEvol} and \citet{Nesvorny2002Trojans}.  With regards to the large Trojans, all particles of (1437) Diomedes escape the L$_4$ swarm by the end of our simulations. 

\begin{figure*}
	\centering
	\subfloat{{\includegraphics[width=0.45\textwidth]{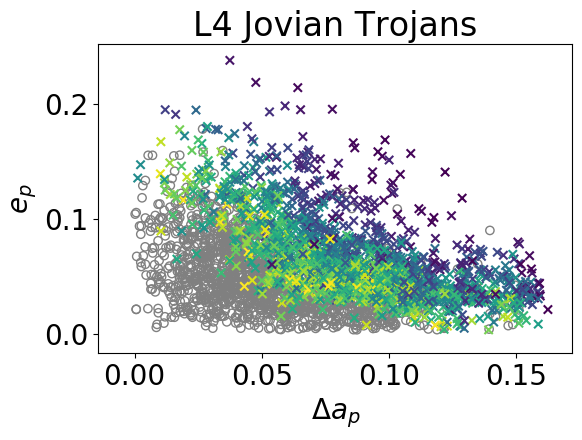} }}%
    \qquad
    \subfloat{{\includegraphics[width=0.45\textwidth]{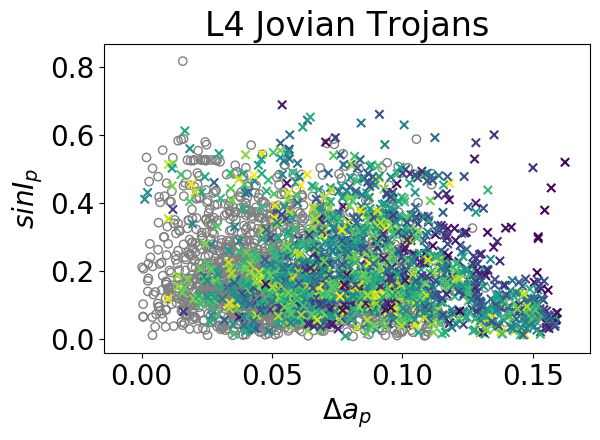} }}%
    \qquad
    \subfloat{{\includegraphics[width=0.45\textwidth]{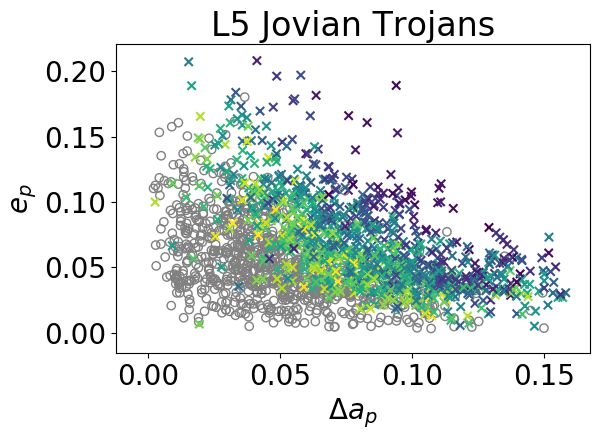} }}%
    \qquad
    \subfloat{{\includegraphics[width=0.45\textwidth]{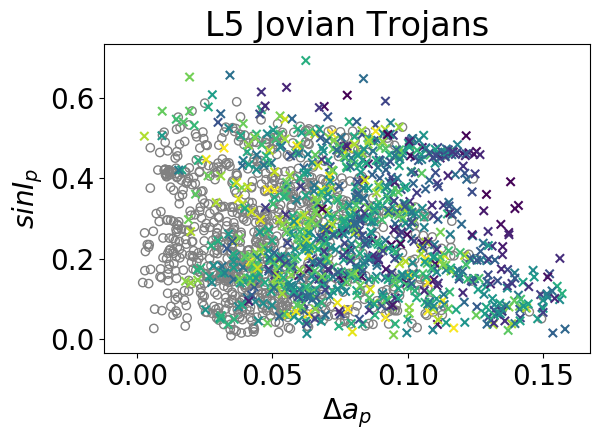} }}%
    \qquad
    \subfloat{{\includegraphics[width=0.9\textwidth]{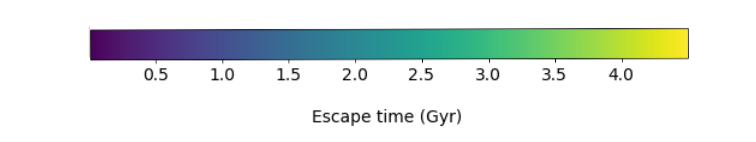} }}%
    \caption{Escape analysis of Jovian Trojans in the L$_4$ and L$_5$ swarms simulated over 4.5 Gyr. Proper elements, semi-major axis ($\Delta a_p$), eccentricity ($e_p$) and sine inclination ($sin I_p$) are taken from the AstDys database \citep{Knezevic2017AstDysTrojans}. \textcolor{gray}{o} indicates objects that are stable over the simulated time frame. X show objects that have at least one particle escaping the population, with their mean respective escape times indicated by colour.}
    \label{Fig:L4L5AvsEvsI}
\end{figure*}

The timing of the reference particle escapes are shown in Fig. \ref{Fig:L4L5AvsEvsI}. With larger changes in semi-major axis ($\Delta a_p$) and eccentricity ($e_p$), there is an increase in the instability. Proper inclination ($sin-i_p$) appears to have little effect on the general instability of the particles. This general trend is consistent with other studies \citep{Nesvorny2002SaturnTrojanHypothetical, Tsiganis2005ChaosJupTrojans, DiSisto2014JupTrojanModels, DiSisto2019TrojanEscapes}. With the inclusion of the timing of escape, we show that there is a gradient to the instability trends, particularly in the $\Delta a_p$ to $e_p$ relationship. This is in a similar unstable parameter space to that identified in \citet{Nesvorny2002SaturnTrojanHypothetical}.

\subsection{Escape Analysis}
\label{SubSec:EscapeAnalysis}
\begin{figure*}
	\centering
    \includegraphics[width=\textwidth]{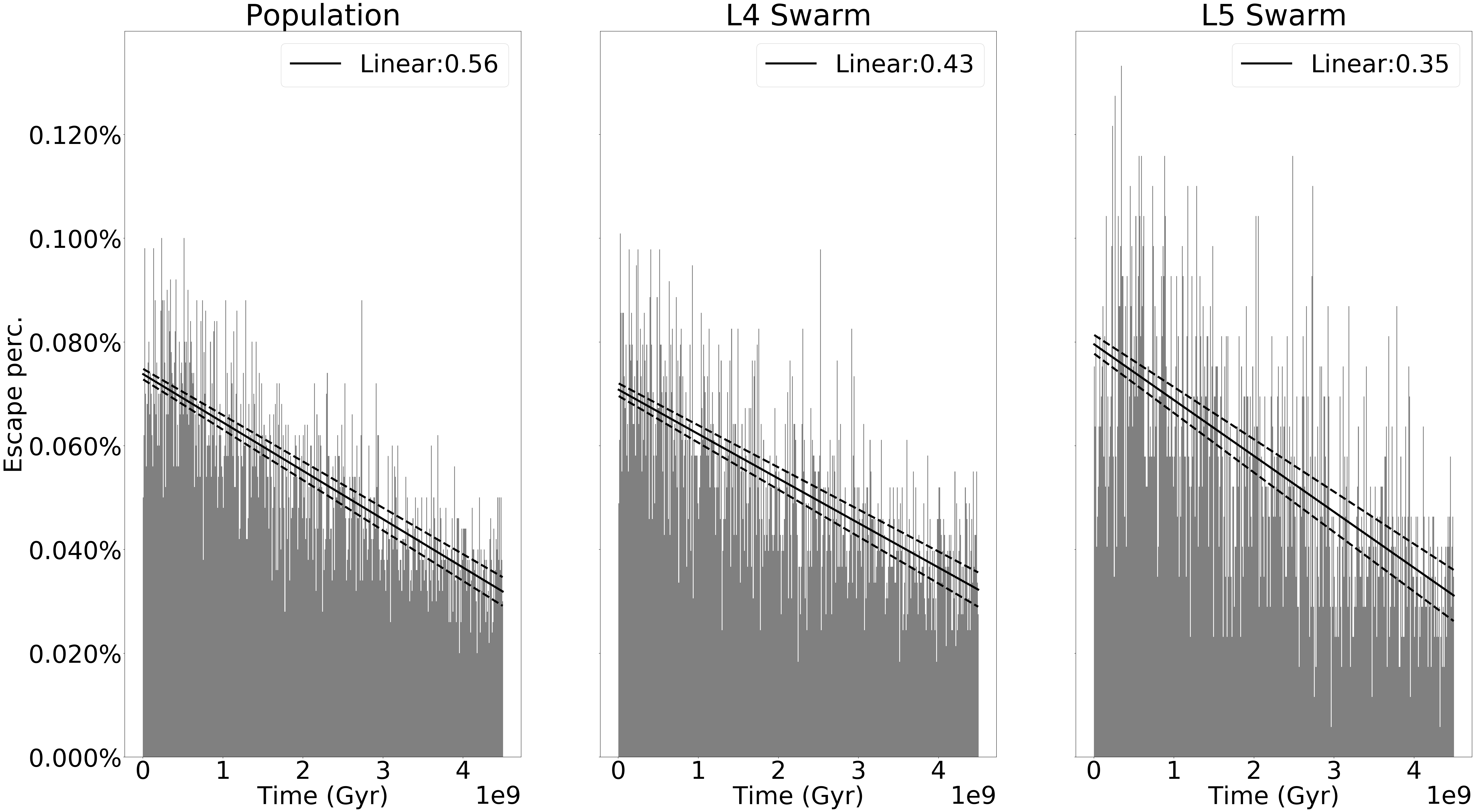} %
    \caption{Histograms of escape percentages of the contemporary number, per $1\e{7}$ years, of a pool of Jovian Trojan particles, in the combined population, L$_4$ and L$_5$ swarms. Lines are linear best fit along with associated $R^2$ values. Dotted lines are $1\sigma$ errors.}
    \label{Fig:JTEscapesHistograms}
\end{figure*}

During our $4.5\e{9}$ year simulations, we track the timing of any particles that escape the Jovian Trojan population. As the orbital elements of our test particles are recorded at intervals of $1\e{5}$ years, the escape times are only accurate to that resolution. For this analysis we pool our results for all test particles considered in this work, including the reference object and each of the eight $1\sigma$ clones, as independent objects. This gives statistical robustness to the analysis. A histogram of the escape percentages for the population as a whole, and each of the L$_4$ and L$_5$ swarms is presented in Fig. \ref{Fig:JTEscapesHistograms}. 

We create linear regression equations to the escape percentages as a function of time, independently for the combined population, and for the L$_4$/L$_5$ swarms. These equations, along with their associated coefficients of determination ($R^2$) and $1\sigma$ errors are presented in Fig. \ref{Fig:JTEscapesHistograms}. These linear fits are shown in equations \ref{Equ:LinearSwarm} for the population, equation \ref{Equ:LinearL$_4$} for the L$_4$ swarm, and equation \ref{Equ:LinearL$_5$} for the L$_5$. In these equations, the escape percentages ($y$) are per $1\e7$ years ($x$) of the contemporary size of the population (equation \ref{Equ:LinearSwarm}) and each individual swarms (equations \ref{Equ:LinearL$_4$}-\ref{Equ:LinearL$_5$}). These equations are similar, once the bins are taken into account, to those found by \citet{DiSisto2019TrojanEscapes}, validating our results. 

\begin{equation}
    y_{pop} = -9.328\e{-14} x + 0.0007384
    \label{Equ:LinearSwarm}
\end{equation}

\begin{equation}
    y_{L4} = -8.581\e{-14} x + 0.0007085
    \label{Equ:LinearL$_4$}
\end{equation}

\begin{equation}
    y_{L5} = -1.078\e{-14} x + 0.000796
    \label{Equ:LinearL$_5$}
\end{equation}

\begin{figure}
	\centering
    \includegraphics[width=\columnwidth]{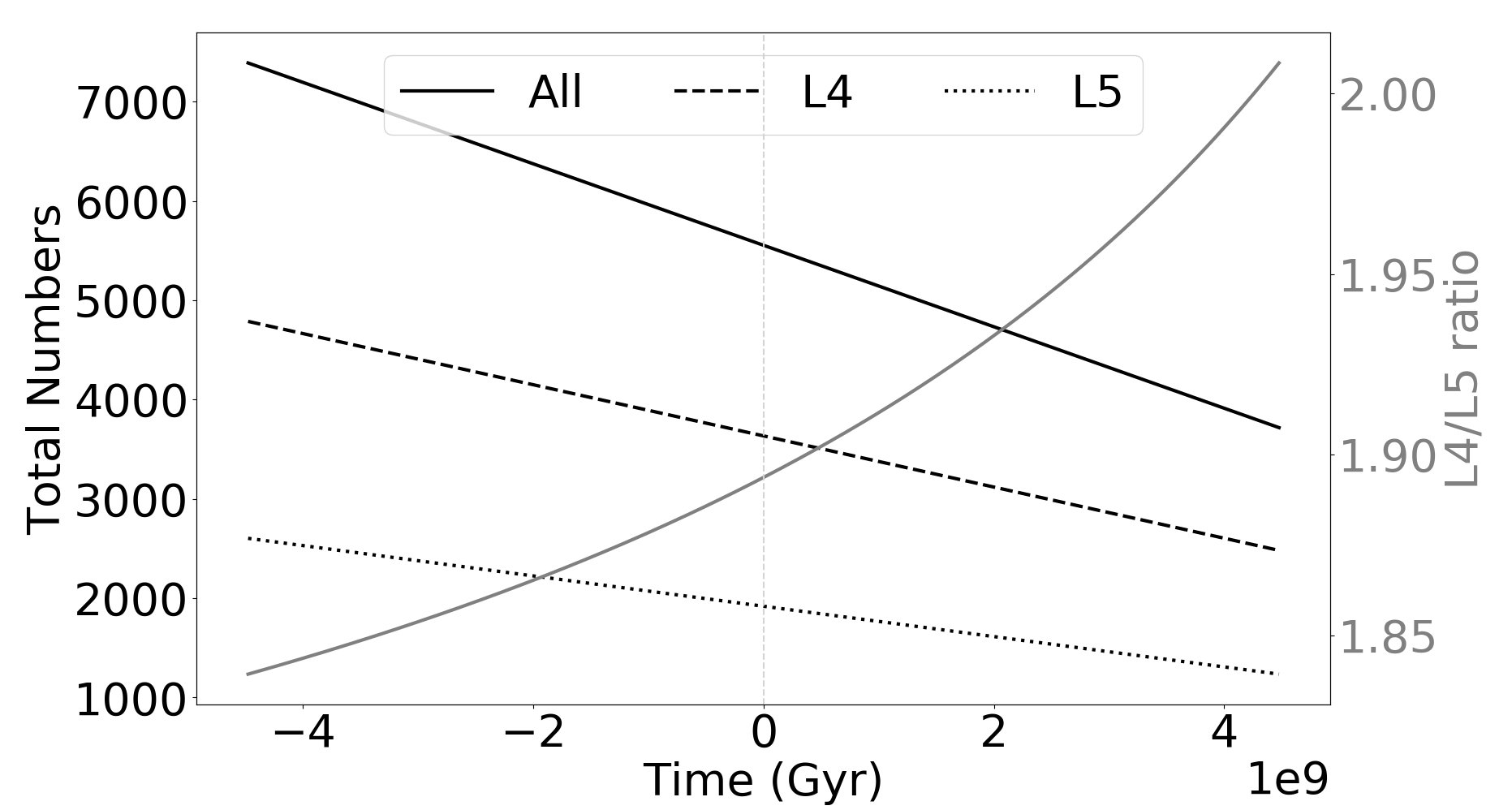}
    \caption{Number of objects, calculated from the contemporary total population (solid line), L$_4$ (dashed line) and L$_5$ (dotted line) Jovian Trojan swarms, as a function of time, with 0 time being the present. Right axis shows changing ratio (gray line) between L$_4$ and L$_5$ swarms. Plotted from equations discussed in section \ref{SubSec:EscapeAnalysis}}
    \label{Fig:L$_4$L$_5$escapeTotalRatio}
\end{figure}

Using linear equations \ref{Equ:LinearSwarm}-\ref{Equ:LinearL$_5$} we can calculate the predicted original size of the Jovian population and L$_4$/L$_5$ swarms, see Fig \ref{Fig:L$_4$L$_5$escapeTotalRatio}, under the assumption that the historical decay of the Trojan population proceeded in the same manner as we see in our simulations. Though the known Jovian Trojan size-frequency distribution, Fig. \ref{Fig:TrojansSFD}, is only complete to a fraction of the theoretical size, we can still make predictions of the number of objects, placing constraints on their formation and capture. The original population, based on the integration of equation \ref{Equ:LinearSwarm}, is approximately $1.332\pm0.004$ times the current population. There is an observed difference in the past size of the L$_4$ and L$_5$ swarms. Due to the difference in their escape rates, the past L$_4$ swarm is predicted to be $1.319\pm0.005$ times larger than the contemporary swarm, while the L$_5$ is $1.358\pm0.008$ times larger. The predicted implantation size, based on modern numbers and the escape rates, are $4792\pm19$ for the L$_4$ and $2606\pm15$ for the L$_5$. This past ratio reduces the current 1.89 numerical asymmetry to $1.84\pm0.003$. This small difference in past/contemporary size ratio does not account for the modern observed numerical asymmetry , as previously noted \citep{Nesvorny2002SaturnTrojanHypothetical, Tsiganis2005ChaosJupTrojans, DiSisto2014JupTrojanModels, DiSisto2019TrojanEscapes}.

The in-situ transport model \citep{Pirani2019PlanetMigrationSSB,Pirani2019TrojanInc} predicts that the initial mass the Jovian Trojan population was three to four times the magnitude of the observed population. Our escape analysis estimates a primordial population size only $1.332\pm0.004$ times larger than today. This is still several orders of magnitude smaller than the most conservative predictions of \citet{Pirani2019PlanetMigrationSSB}. However, it should be noted that our estimates for the initial population are based on the assumption that the current linear decay has remained consistent since the origin of the Trojan population. In the population's youth, it is possible that the decay rate could have been markedly higher, had objects been efficiently captured to the less stable regions of the Trojan population. \citet{Pirani2019TrojanInc} do report on interactions with Saturn affecting Trojans larger inclinations, though this is still insufficient to explain the current escape rate.

\begin{figure}
	\centering
    \includegraphics[width=\columnwidth]{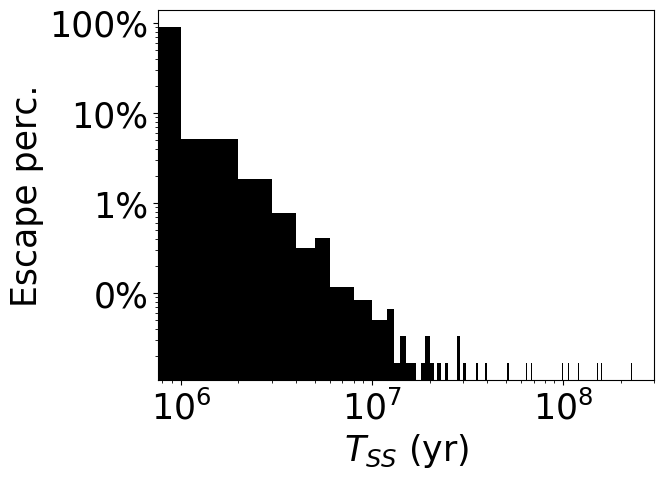} 
    \caption{Histogram ($1\e{6}$ year bins) of time spent in the Solar system prior to ejection ($T_{SS}$), of objects that escape the Jovian Trojan population. Escape percentages are based of nine particles generated for each of 5553 Jovian Trojans.}
    \label{Fig:SSOParticipation}
\end{figure}

\begin{figure}
    \includegraphics[width=\columnwidth]{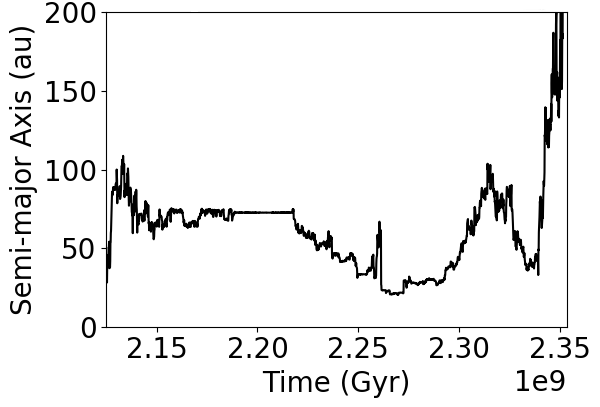}
    \caption{The behaviour of the longest lived escapee, clone 2 of (312627) 2009 TS$_{26}$ in semi-major axis over time. Start time is the point when the particle escapes the L4 Jovian swarm. End time is when the particle escapes the Solar system} 
    \label{Fig:LLEscape}
\end{figure}

The majority of escape particles are eventually ejected from the Solar system, by achieving a heliocentric distance of 1000~au, in the same $1\e{5}$ time-step. This is longer than the expected life time of most Centaurs \citep{Horner2004CentaursI}, particularly those starting on orbits close to that of Jupiter. A fraction of the population escapees, approximately $41.41\%$, stay within the Solar system for a longer period of time, prior to being ejected. This fraction is similar between the L$_4$ and L$_5$ populations, $41.37\%$ and $41.45\%$ respectively. This similarity between swarms is not unexpected, since the chaotic evolution of test particles once they leave the Trojan population would be expected to quickly erase any `memory' of their original orbit. Fig. \ref{Fig:SSOParticipation} shows the length of time that these particles spend in the Solar system, with over $88.58\%$ escaping in the first $1\e{6}$ years, and an additional $6.15\%$ escaping in the next $1.0\e{6}$ years. By $1.0\e{7}$ years, $99.25\%$ of the particles have been ejected. These short lifetimes are consistent with the expected lifetimes of Centaurs \citep{Horner2004CentaursI}. \citet{Horner2012AnchisesThermDynam} show that at least one escaped Jovian Trojan, (1173) Anchises, can participate in the Centaur population before being ejected. Despite this high number of short lived objects, 13 particles survive longer than $3.2\e{7}$ years, the expected lifetime of the longest Centaur \citep{Horner2004CentaursI}. These long lived particles are not unexpected, as \citet{Horner2004CentaursI, Horner2004CentaursII} also reported on several long lived particles. Each of our clone particles have a different reference object. The longest lived particle is clone 2 of (312627) 2009 TS$_{26}$, which lives for $2.286\e{8}$ years, shown in Fig. \ref{Fig:LLEscape}, and represents a typical chaotic pattern for escaped Trojans. 

Less than 10\%, 547 objects, of the Jovian Trojan population has been classified under the Bus-Demeo system \citep{Tholen1984Taxonomy, Bus2002AsteroidTax, Bendjoya2004JTSpectra, Fornasier2004L5TrojanSpec, Fornasier2007VisSpecTrojans, DeMeo2009AsteroidTax,  Carvano2010SDSSAstTax, Grav2012JupTrojanWISE, DeMeo2013SDSSTaxonomy}. The majority, 65.08\%, are considered D-types, with several other minor classes X-type (15.17\%), C-type (12.79\%) and other classes below 5\% (P-type, L-type, S-type, V-type and F-type). The rate at which the three major classes, D-type, X-type and C-type objects escape, 23.00\%, 27.66\% and 24.13\% respectively, is roughly constant with the overall population. Many of the smaller taxonomic classes come from  \citet{Carvano2010SDSSAstTax, Hasselmann2012SDSSTaxonomy}, and have low classification confidence levels. If we reduce the taxonomic data-set to only those in \citet{Carvano2010SDSSAstTax, Hasselmann2012SDSSTaxonomy} with a confidence classification of greater than 50, it reduces the classified Trojans down to 2\% of the population, and only D-Type (79.24\%), X-type (14.15\%) and C-Type (6.6\%) objects. This restriction does not change the escape rates significantly for the D-Types at 23.41\%. The X-types and C-types do increase to 32.59\% and 31.75\% respectively, though these classes suffer from the variances of small number statistics. This classification analysis is something that may merit further study once data becomes available from the Rubin Observatory Legacy Survey of Space and Time (LSST) \citep{Schwamb2018LSSTTrojans, Schwamb2018LSST}, and our escape analysis can then be placed in a wider taxonomic context.

\section{Collisional Families}
\label{Sec:CollisionalFamilies}
In order to further investigate the escapes of collisional family members, we have increased the number of clones simulated to 125 for each of the canonical family members in \citet{Nesvorny2015AsteroidFamsAIV}. This increases the statistical significance of the escape analysis. For comparison purposes, the wider, non-canonical family datasets found by \citet{Broz2011EurybatesFamily} and \citet{Rozehnal2016HektorTaxon} use the original eight clones, as in section \ref{Sec:L4L5Swarms}, and only those objects found in the AstDys database \citep{Knezevic2017AstDysTrojans}. 

The specific numbers of canonical collisional family members that are simulated in this work are shown in Table \ref{Tab:families}, after \citet{Nesvorny2015AsteroidFamsAIV}. Of particular interest is the Eurybates family. This is the largest known family in the Jovian Trojan population, and is discussed separately in section \ref{SubSec:Eurybates}. When all of the particles are considered independently, $f_{EscP}$ and $f_{VEscP}$ in Table \ref{Tab:families}, the percentage that escape is similar to the escape rate of the reference particles ($f_{EscR}$ and $f_{VEscR}$ in Table \ref{Tab:families}). This is comparable to the trends seen in the overall swarms, see section \ref{Sec:L4L5Swarms}. 

\begin{table*}
	\centering
	\caption{Escaping collisional family members; $n$: number of objects in each canonical collisional family \citep{Nesvorny2015AsteroidFamsAIV}; the  $n_{EscR}$: number of reference particles that escape:  $f_{EscR}$: numerical percentage of reference particles that escape; $f_{VEscR}$: volumetric percentage of reference particles that escape; $f_{EscP}$: numerical percentage Trojan particle pool, Reference and 125 $1\sigma$ clones, that escape; $f_{VEscP}$: volumetric percentage Trojan particle pool, Reference and 125 $1\sigma$ clones, that escape}
	\label{Tab:families}
\begin{tabular}{lcccccc}
\hline
               & $n$ & $n_{EscR}$ & $f_{EscR}$ & $f_{VEscR}$ & $f_{EscP}$ & $f_{VEscP}$ \\
               \hline
\textbf{L$_4$ Families }   &     &            &            &             &             &              \\
Eurybates (1)  & 218 & 43         & 19.72\%    & 7.43\%      & 19.59\%     & 8.05\%      \\
Hektor (2)     & 12  & 2          & 16.66\%    & 0.06\%      & 11.99\%     & 28.53\%      \\
1996 RJ (3)    & 7   & 0          & 0.00\%     & 0.00\%      & 0.00\%      & 0.00\%       \\
Arkesilaos (4) & 37  & 1          & 2.70\%     & 1.13\%      & 3.09\%      & 3.47\%       \\
\hline
\textbf{L$_5$ Families}    &     &            &            &             &             &              \\
Ennomos (5)    & 30  & 15         & 50.00\%    & 66.39\%     & 34.29\%     & 17.47\%      \\
2001 UV$_{209}$ (6) & 13  & 0          & 0.00\%     & 0.00\%      & 0.00\%      & 0.00\%       \\
\hline
Total          & 317 & 61         & 19.24\%    & 12.45\%     & 17.67\%     & 24.75\%        
\end{tabular}%
\end{table*}

In general terms, the members of known collisional families within our integrations show lower escape percentages than the total of the swarms. This is due to the fact that the majority of the known collisional families are located in the more stable regions of the delta semi-major axis, eccentricity and sin $i$ parameter space, as shown in Fig. \ref{Fig:L$_4$FamiliesAEI} and Fig. \ref{Fig:L$_5$FamiliesAEI}. 

There are also potentially a significant number of undetected family members \citep{Yoshida2008JovTrojanSize, Vinogradova2015JupTrojanMass} in the Jovian Trojan population. The numerical escape percentages may increase as a larger number of objects are discovered by new surveys, such as the Rubin Observatory Legacy Survey of Space and Time (LSST) \citep{Schwamb2018LSST}, which is expected to commence science operations in 2023. As these new objects are discovered, their allocation to collisional families and long-term stabilities will need to be investigated. 

\subsection{L4 collisional Families}
\begin{figure*}
	\centering
    \subfloat{{\includegraphics[width=0.45\textwidth]{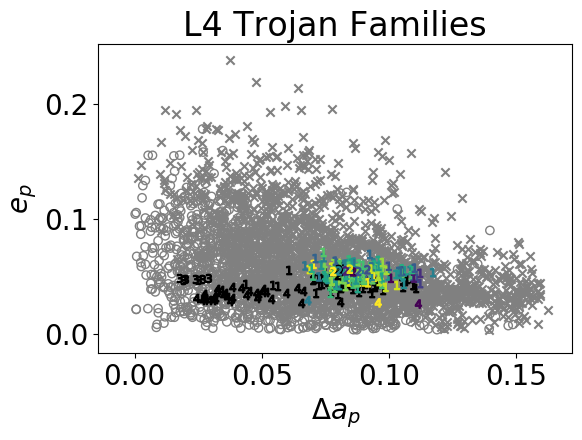} }}%
    \qquad
    \subfloat{{\includegraphics[width=0.45\textwidth]{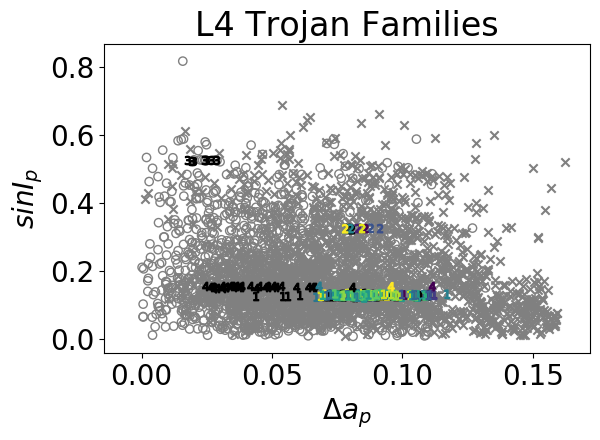} }}%
    \qquad
    \subfloat{{\includegraphics[width=0.9\textwidth]{ColourBarViridis.png} }}%
    \caption{Escape analysis of collisional family members located in the L$_4$ Jovian Trojan swarm simulated for $4.5\e{9}$ years. Shown are the instabilities of the reference object. Proper elements, semi-major axis ($\Delta a_p$), eccentricity ($e_p$) and sine inclination ($sin I_p$), are taken from the AstDys database \citep{Knezevic2017AstDysTrojans}. \textcolor{gray}{o} indicates objects that are stable over the simulated timeframe. \textcolor{gray}{x} are unstable background objects. Family membership: Eurybates (1), Hektor (2), 1996 RJ (3), Arkesilaos (4). Black numbers are stable, with colours showing mean particle escape time.}
    \label{Fig:L$_4$FamiliesAEI}
\end{figure*}

In the L$_4$ swarm, shown in Fig. \ref{Fig:L$_4$FamiliesAEI} a total four families have been identified. The largest L$_4$ cluster, the Eurybates family is discussed in section \ref{SubSec:Eurybates}.

\subsubsection{Eurybates family}
\label{SubSec:Eurybates}
The Eurybates family is the largest and most consistently identified \citep{Broz2011EurybatesFamily, Nesvorny2015AsteroidFamsAIV,  Vinogradova2015TrjoanFamilies} collisional cluster in the Trojan population. The largest fragment of the family, (5348) Eurybates, is also the target of future visitation by the \textit{Lucy} spacecraft in 2027 \citep{Levison2017Lucy}. In our simulations, we consider the canonical 218 identified members of the family \citep{Nesvorny2015AsteroidFamsAIV}. From the 310 members identified by \citet{Broz2011EurybatesFamily}, 293 are in the AstDys database. In the canconcial members, there is a 19.59\% escape percentage for the particle pool. If we consider the larger set identified by \citep{Broz2011EurybatesFamily}, this escape percentage only decreases slightly to 19.07\%.

\begin{figure}
	\centering
    \includegraphics[width=1\columnwidth]{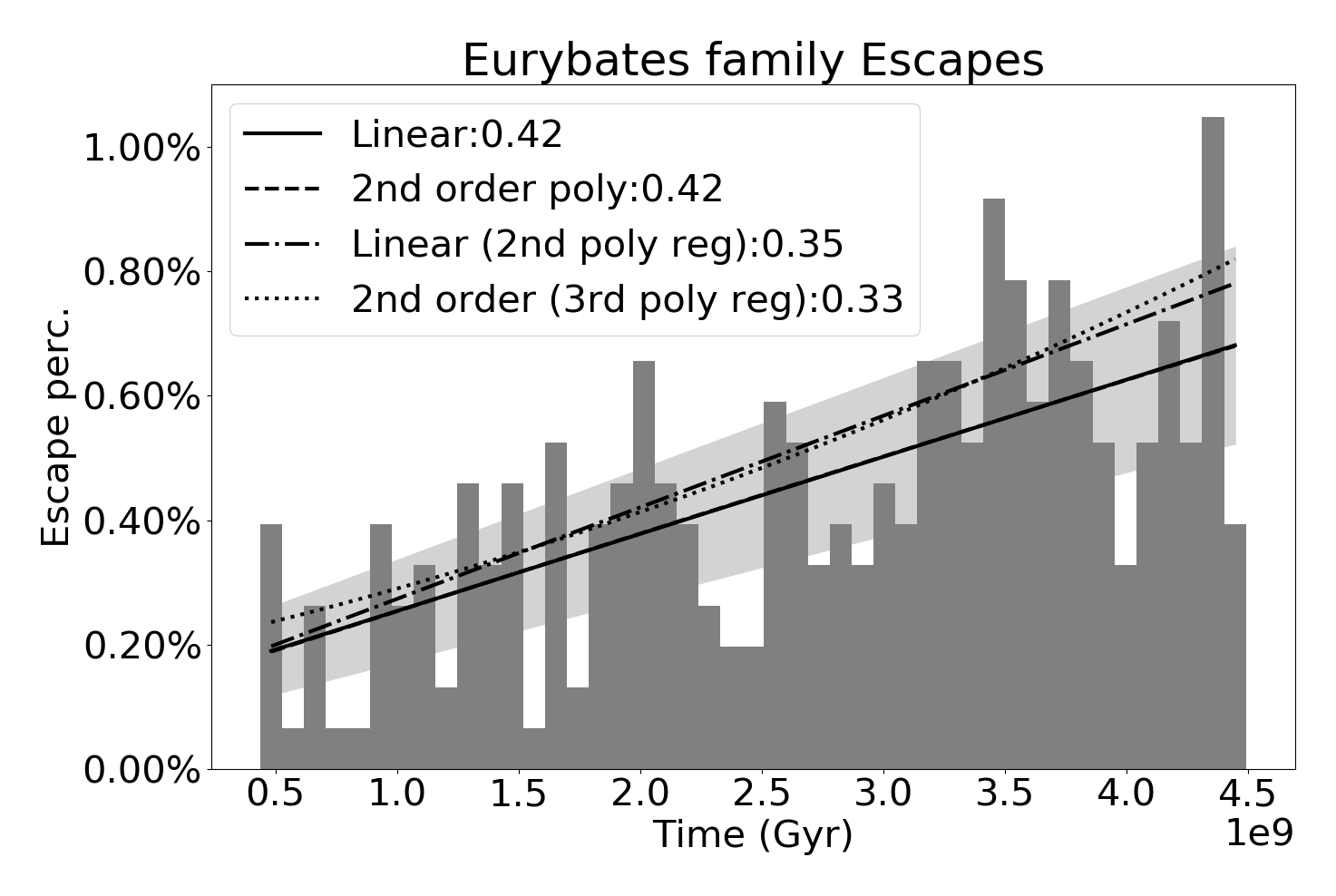} %
    \caption{Histogram ($1\e{8}$ year bins) of escapes from the Eurybates collisional family. Lines indicate best fit analysis scaled to the histogram bins, with $R^2$ scores for linear fit (solid, with  Light grey shading indicating $1\sigma$ error) and second degree poylnomial (dashed) lines. Fits are also shown from the results of linear regression analysis on second (dot-dashed) and third order polynomial (doted) generated from a cumulative histogram.}
    \label{Fig:EurybatesEscapeAnalysis}
\end{figure}

As was seen in the L$_4$ swarm (Fig. \ref{Sec:L4L5Swarms}), there is a gradient to the escapes, with larger changes in semi-major axis ($\Delta a_p$) and eccentricity ($e_p$), causing particles to escape the swarm sooner. Contrary to the overall decreasing escape rates seen in the L$_4$ swarm, we found the escape rate of the Eurybatyes family to be increasing with time, as can be seen in Fig. \ref{Fig:EurybatesEscapeAnalysis}. A possible explanation for this is the ongoing diffusion of family members into less stable parameter space, as they disperse chaotically from the initial location of the breakup event. Such dispersion can be seen in main belt families \citep{Milani1992ProperElements, Bottke2005SFDCollisions, Broz2013EosFamily, Aljbaae2019GefionFamily}, with members gradually diffusing into Jovian resonances and being ejected from the main belt. Future simulations of a synthetic Eurybates family would be required to confirm this, and are beyond the scope of this paper. 

As with the L$_4$ swarm escape analysis, a standard linear regression offers the most reliable fit for the data. We did attempt to create a second order polynomial, along with using cumulative linear and polynomial regression to improve the fit in this case, though as Fig. \ref{Fig:EurybatesEscapeAnalysis} demonstrates, this did not improve the coefficient of determination. The coefficient of determination for the linear fit ($R^2$ = 0.42) is similar to the L$_4$ swarm, due to number of particles being considered being an order of magnitude smaller. We attempted to take account for this by using an order of magnitude larger bins to increase the number of ejections per bin to a reasonable number. The y-intercept of this linear equation, which represents the time at which the escape rate from the Eurybates family equals zero, might be considered to be an indication of the age of the family. If such a conclusion is reasonable, our data would place the family formation event some $1.045\pm 0.364\e9$ years ago. This age is presented as a minimum age, though preliminary simulations of a synthetic Eurybates family \citep{Holt2019SythEurybates} indicate that the observed dynamical situation could be achieved within $1\e{5}$ years. As previously stated, the two other methods of collisional family age estimation, high precision reverse integration \citep{Nesvorny2002AsteroidBreakup} and Yarkvosky `V' \citep{Milani2017AsteroidFamAges} are inappropriate for the Trojan families. Using a small number of synthetic members, \citet{Broz2011EurybatesFamily} also calculated a wide time range, 1Gyr--4Gyr, for the family creation event. Our age is therefore one of the first estimations that give a reasonable order of magnitude age and constrained range for the Eurybates family. As larger numbers of family members are identified, a re-investigation should improve the statistical reliability of this analysis.

\begin{figure*}

	\centering
    \subfloat{{\includegraphics[width=0.45\textwidth]{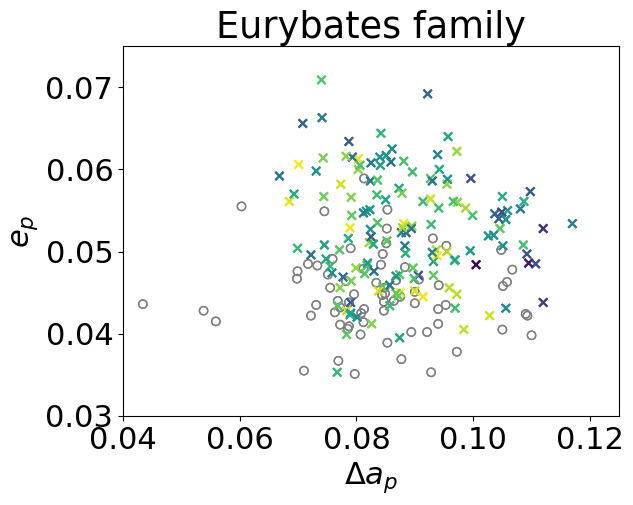} }}%
    \qquad
    \subfloat{{\includegraphics[width=0.45\textwidth]{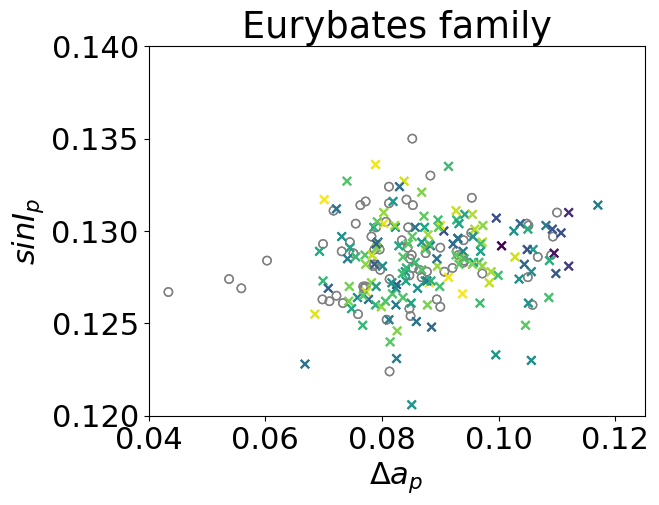} }}%
    \qquad
    \subfloat{{\includegraphics[width=0.9\textwidth]{ColourBarViridis.png} }}%
    \caption{Escape analysis of the canonical Eurybates collisional family members identified in \citet{Nesvorny2015AsteroidFamsAIV}, simulated for $4.5\e{9}$ years. Shown are the mean escape time of 126 particles for the object (coloured x). Proper elements, semi-major axis ($\Delta a_p$), eccentricity ($e_p$) and sine inclination ($sin I_p$), are taken from the AstDys database \citep{Knezevic2017AstDysTrojans}. \textcolor{gray}{o} indicates objects that are stable over the simulated time frame.}
    \label{Fig:EurybatesAEI}
\end{figure*}

\subsubsection{Hektor family}
\citet{Rozehnal2016HektorTaxon} identified 90 objects in this family, using the Random box method. We use the canonical twelve objects from \citet{Nesvorny2015AsteroidFamsAIV}, and note where there could possibly be a different escape rate. The family is characterised by a moderate $\Delta a_p$ and $e_p$, with a comparatively high $sinI_p$. The parent body, (624) Hektor has been classified under the Bus-Demeo spectral taxonomy \citep{DeMeo2009AsteroidTax} as a D-type asteroid \citep{Emery2006SpectraJovianTrojan,Emery2011JovTrojanIRComp,Rozehnal2016HektorTaxon}. It is also a contact binary, with a confirmed satellite \citep{Marchis2014Hektor}. The canonical Hektor family has a low escape rate, with only two reference particles from the family eventually escaping the swarm. One of these is the reference particle of (624) Hektor itself, which also has a 28.8\% particle escape rate. These particles account for the large volume of escapes, nearly double that of the numerical escape fraction. Unfortunately, the small number of identified members of the Hektor family, twelve known objects, means that a statistical analysis of these results would prove problematic. Using the larger number of clones, we can assign a numerical escape percentage of 12\%.  If the wider numbers, 77 objects from \citet{Rozehnal2016HektorTaxon} are used, then 18.18\% of particles escape. 

\subsubsection{1996 RJ family}
The compact 1996 RJ family has a small $\Delta a_p$ and $e_p$. This places it firmly within the predicted stability region from \citet{Nesvorny2002SaturnTrojanHypothetical}. The high inclinations of the family members do not seem to have an effect on their stability. Our results show that this family is completely stable, with no escapes. Those members from \citet{Rozehnal2016HektorTaxon} are also stable, except for the single particle, clone 6 of (195104) 2002 CN$_{130}$. This particular object has a higher $\Delta a_p$ than the rest of the family, and is a probable outlier. 

\subsubsection{Arkesilaos family}
This is a medium sized family, with 37 cannonical members. It is confirmed by \citet{Vinogradova2015TrjoanFamilies}, though they use (2148) Epeios as the main object and have a larger number of members (130). \citet{Rozehnal2016HektorTaxon} chose (20961) Arkesilaos as the primary objects due to consistency at the center of the family parameter space, even at low cut-off velocities. The family has a wide distribution of $\Delta a_p$ values and a compact range of $e_p$ and $sinI_p$ values. Predictably, the family is stable with three small outliers that escape. (356237) 2009 SA$_328$ is the most unstable, with 72\% of the particles escaping. This is due to its high $\Delta a_p$, placing it in the unstable parameter space. (394808) 2008 RV$_{124}$ and (20961) Arkesilaos also have some particles escape, but only 28.9\% and 14.4\% respectively. The escape fraction of the family only changes slightly to 2.24\%, considering the additional members identified by \citet{Rozehnal2016HektorTaxon}. The small escape percentages of this family preclude any additional statistical analysis.

\subsection{L5 Collisional families}
\begin{figure*}
	\centering
    \subfloat{{\includegraphics[width=0.45\textwidth]{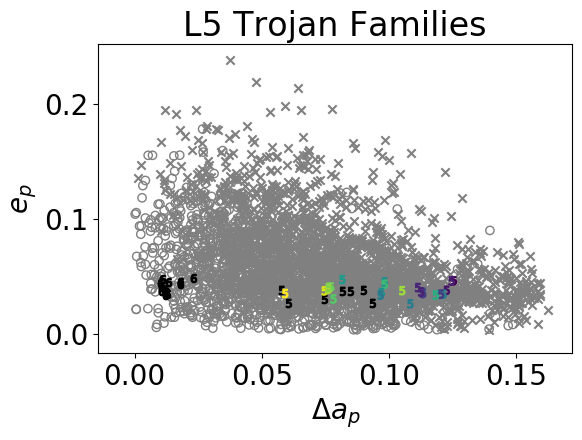} }}%
    \qquad
    \subfloat{{\includegraphics[width=0.45\textwidth]{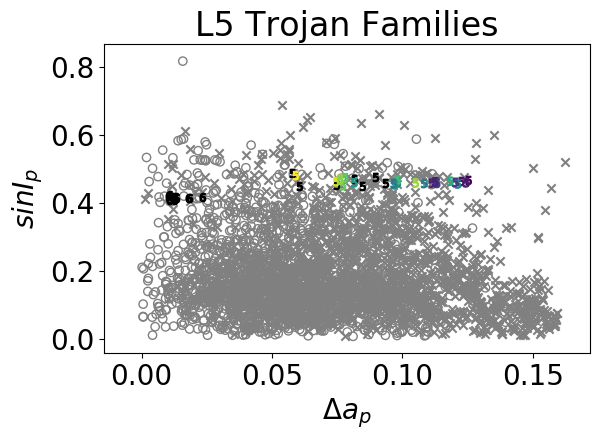} }}%
    \qquad
    \subfloat{{\includegraphics[width=0.9\textwidth]{ColourBarViridis.png} }}%
    \caption{Escape analysis of collisional family members located in the L$_5$ Jovian Trojan Swarm simulated for 4.5e9 years. Proper elements, delta semi-major axis ($\Delta a_p$), eccentricity ($e_p$) and sine inclination ($sin I_p$) are taken from the AstDys database \citep{Knezevic2017AstDysTrojans}. \textcolor{gray}{o} indicates objects that are stable over the simulated time frame. \textcolor{gray}{x} are unstable background objects. Numbers indicate collisional family membership: Ennomos (5), 2001 UV$_{209}$ (6). Black numbers are stable, with colours showing mean escape time of 126 particles for the object.}
    \label{Fig:L$_5$FamiliesAEI}
\end{figure*}

Within the L5 swarm, there are only two identified collisional families \citep{Nesvorny2015AsteroidFamsAIV}, the Ennomos and 2001 UV$_{209}$ families. Contrary to \citet{Rozehnal2016HektorTaxon} and the canonical \citet{Nesvorny2015AsteroidFamsAIV}, \citet{Vinogradova2015TrjoanFamilies} do not consider either of the families valid, though they note that there is some clustering around the largest members. We show the escape times of the L5 families in Fig. \ref{Fig:L$_5$FamiliesAEI}.

 \subsubsection{Ennomos family}
The most unstable cluster in the L5 swarm is the Ennomos family. This is a medium sized cluster, with 30 identified objects in \citet{Nesvorny2015AsteroidFamsAIV}. There are a larger number of objects, 104, of which 85 are in the Astdys database, identified by \citep{Rozehnal2016HektorTaxon}. The family members have relatively high $\Delta a_p$ and $sinI_P$, with low $e$, placing them on the edge of the stable parameter space. Consequently, a large fraction of Ennomos family members, 50\% of reference particles, escape the swarm. When considering just the reference particles, $66.66\%$ of the volume escape during our simulations. This is due to the reference particle and a low number of clones (14.28\%) of (1867) Deiphobus, a 59km object, escaping the L$_5$ swarm. In the more statistically robust particle pool, the escape percentage by volume drops to 17.47\%. This family is characterized by its high inclination and delta semi-major axis, so a high amount of instability is not unexpected. In this family, there are three members, (48373) Gorgythion, (381987) 2010 HZ$_{21}$ and (287454) 2002 YX$_7$ where all particles escape. This is unsurprising, as (48373) Gorgythion has the largest proper $\Delta a_p$ and $e_p$ of the family. In addition to these three, six objects have over 50\% of their particles escape.  Including the larger number of members from \citet{Rozehnal2016HektorTaxon}, decreases the escape rate to 23.14\%, closer to the overall L5 rate.

As in section \ref{SubSec:Eurybates}, we attempted regression analysis to ascertain the age of this family. \citet{Broz2011EurybatesFamily} estimate the age of the family to be approximately 1--2 Gya. Similar to the L5 swarm and unlike the Eurybates family, the slope of the linear regression analysis is negative, though fairly flat ($-1.62\e{-12}$). The $R^2$ score is only 0.13, so until additional family members are identified, these are only preliminary indications. 

\subsubsection{2001 UV$_{209}$}
This small family, with thirteen canonical members, is located well within the stable $\Delta a_p$ - $e$ parameter space. It is then not unexpected that the 2001 UV$_{209}$ family members are stable in our simulations. Considering the expanded 36 objects identified by \citet{Rozehnal2016HektorTaxon}, this jumps to 13.89\%. These unstable members are not considered valid by \citet{Nesvorny2015AsteroidFamsAIV}, and with higher $\Delta a_p$ are probable background objects, rather than members of the family.

\section{Conclusions}
\label{Sec:Conclusions}
The Jovian Trojans are a fascinating collection of objects, remnants of the early stages of the Solar system's formation. In this work, we present the results of detailed $n-$body simulations of the known Jovian Trojan population, using nearly double the number of objects of the previous largest study \citep{DiSisto2014JupTrojanModels, DiSisto2019TrojanEscapes}. We simulate the orbital evolution of a population of 49,977 massless test particles, nine particles for each of the 5553 known Jovian Trojans, for a period of $4.5\e{9}$ years into the future, under the gravitational influence of the Sun and the four giant planets. Our simulations reveal that, the populations of both the L$_4$ and L$_5$ swarms are predominately stable, however a significant number of objects from both swarms can escape over the lifetime of the Solar system. In the case of the leading L$_4$ swarm, we find that 23.35\% of objects  escape, by volume. Similarly, only 24.89\% escape the trailing L$_5$ swarm. Overall, 23.95\% by volume of all test particles simulated in this work escape the Jovian population. As discussed by other authors \citep{Nesvorny2002SaturnTrojanHypothetical, Tsiganis2005ChaosJupTrojans, Nesvorny2013TrojanCaptureJJ, DiSisto2014JupTrojanModels, DiSisto2019TrojanEscapes}, we find that the escape rates can not explain the current observed asymmetry between the two swarms. This supports the conclusion that the observed asymmetry between the L$_4$ and L$_5$ swarms are the result of their initial capture implantation \citep{Nesvorny2013TrojanCaptureJJ, Pirani2019PlanetMigrationSSB}. 

The escape rates of objects from the two Trojan swarms are in accordance with the idea that the Jovian Trojans act as a source of material to the other small Solar system body populations, as noted in  \citet{Levison1997JupTrojanEvol,DiSisto2014JupTrojanModels,DiSisto2019TrojanEscapes}, particularly with regards to the Centaurs \citep{Horner2004CentaursI,Horner2012AnchisesThermDynam}. The majority of escaped Trojans, 58.63\%, are ejected from the population and the Solar system within a single $1\e{5}$ year timestep. For those that remain in the Solar system, 99.25\% are ejected by $1\e{7}$ years, after joining the Centaur population. 

In the Jovian Trojan swarms, a total of six collisional families have been identified to date \citep{Nesvorny2015AsteroidFamsAIV}, with four in the L$_4$ swarm and two located around L$_5$. We find that three of the families are highly dynamically stable, with no particles escaping the Trojan population through the course of our integrations (the 1996 RJ,  Arkesilaos and 2001 UV$_{209}$ families). 
Two other collisional groups, the L4 Hektor and L5 Ennomos families did have members that escape. These unstable families all have a small number of known members, which limits our ability to study their stability further in this work. 
The largest known Trojan family, the Eurybates L4 family, has a smaller escape rate than the overall population. Contrary to the escape trends in the population, however, the escape rate of the Eurybates family is found to increase with time in our simulations. This might point to the diffusion of its members into unstable parameter space as they evolve away from the location of the family's creation. From this escape rate, we can obtain an estimate of the age of the Eurybates family on the order of $1.045\pm 0.364\e9$ years. 

In the future, as more members of the Jovian Trojans and their taxonomic groupings are identified, it will be interesting to see whether these dynamical methods can be used to help constrain the ages of the smaller clusters. If this is possible, such results would shed light on the variability of the collision rates within the Jovian Trojan swarms. The results we present in this paper, and these potential future works, highlight the impotence of the Jovian Trojan swarms, their taxonomic groups and collisional families, to understanding the history of the Solar system.

\section*{Acknowledgements}
This research was in part supported by the University of Southern Queensland's Strategic Research Initiative program. TRH was supported by the Australian Government Research Training Program Scholarship. This work makes use of the Anaconda Python software environment \citep{Anaconda240}. We thank Hal Levison for discussions on the paper and for providing previously unpublished data. We also thank Douglas Hamilton and Romina Di Sisto (as a reviewer) for providing comments and insights on this paper. This research has made use of NASA Astrophysics Data System Bibliographic Services. 




\bibliographystyle{mnras}
\bibliography{library}







\bsp	
\label{lastpage}
\end{document}